# Dynamic plasmonic colour display

*Xiaoyang Duan[1,2], Simon Kamin[1,2], & Na Liu[1,2]\**

[1]Max Planck Institute for Intelligent Systems, Heisenbergstrasse 3, 70569 Stuttgart, Germany
[2]Kirchhoff Institute for Physics, University of Heidelberg, Im Neuenheimer Feld 227, 69120 Heidelberg, Germany
\*E-mail: laura.liu@is.mpg.de

**Plasmonic colour printing based on engineered metasurfaces has revolutionized colour display science due to its unprecedented subwavelength resolution and high-density optical data storage. However, advanced plasmonic displays with novel functionalities including dynamic multicolour printing, animations, and highly secure encryption have remained in their infancy. Here we demonstrate a dynamic plasmonic colour display technique which enables all the aforementioned functionalities using catalytic magnesium (Mg) metasurfaces. Controlled hydrogenation and dehydrogenation of the constituent Mg nanoparticles, which serve as dynamic pixels, allow for plasmonic colour printing, tuning, erasing, and restoring. Different dynamic pixels feature distinct colour transformation kinetics, enabling the first plasmonic animations. Through smart material processing, information encoded on selected pixels, which are indiscernible to both optical and scanning electron microscopies, can only be read out using hydrogen as decoding key, suggesting a new generation of information encryption and anti-counterfeiting applications.**

Plasmonic colour generation based on metallic metasurfaces has suggested an exciting new route towards producing colour images with resolution far beyond the limit of current display technologies[1–8]. Although in its infancy, significant advances in plasmonic colour generation have been accomplished including the realization of colour displays at the optical diffraction limit[1–3], scalable full-colour chromotropic printing[4–7], actively tunable plasmonic surfaces[8–11], plasmonic colour laser printing[12], *etc*. Apart from these primary applications, plasmonic colour display technology also features great potentials in stereoscopic imaging[13], high-density data storage[13,14], and anticounterfeiting applications[4,15]. Nevertheless, the fast development of this field inevitably calls for plasmonic displays with advanced functionalities, for example, dynamic multicolour printing, animations, and highly secure encryption.

In this work, we demonstrate a novel dynamic plasmonic display technique based on catalytic Mg metasurfaces. Different from other hydrogen-storage metals such as palladium (Pd)[16] and yttrium[17], which are associated with poor optical response, Mg exhibits excellent plasmonic properties at high frequencies[18–21]. For example, Mg nanostructures have been used for chiral sensing in the UV spectral range and for hydrogen sensing in the visible spectral range[20,21]. Most importantly, the unique hydrogenation/dehydrogenation kinetics of Mg nanoparticles is ideally suited for creating dynamic plasmonic systems. Here, we demonstrate successively plasmonic microprint displays with high reversibility, the first plasmonic animations with subwavelength resolution, and the first highly secure plasmonic encryption. Our work will stimulate fascinating colour display applications utilizing plasmonic dynamic pixels with nanoscale controllability.

**Results**

The concept of our dynamic plasmonic colour display technique is illustrated in Fig. 1a. The dynamic plasmonic pixels comprise Mg nanoparticles, which are sandwiched between titanium (Ti)/Pd capping layers and a Ti adhesion layer. These particles reside on a silicon dioxide film (100 nm) supported by a silicon substrate. In order to render a palette of colours with a broad range, stepwise tuning of the lateral particle size $s$ and the interparticle distance $d$ is employed. Each square (10 μm), which presents a characteristic colour, contains Mg particles that are arranged in a lattice with periodicity of $s+d$.

The brilliant colours from the palette can be revealed using reflection bright-field microscopy as shown in Fig. 1b. The charge-coupled device (CCD) image of the palette is captured using a $20\times$ objective with a numerical aperture (NA) of 0.4. The spectral analysis of the representative colour squares (indicated by the individual horizontal and vertical boxes in Fig. 1b) is carried out and presented in Fig. 1c. The scanning electron microscopy (SEM) images of the corresponding structures are also included in the same figure. Enlarged SEM images of the representative structures can be found in Supplementary Fig. S1. When $d$ increases successively ($s = 150$ nm), a wide range of colours are observed (see the left plot in Fig. 1c). The reflectance peak exhibits evident red-shifts (see the red-dashed line) with increasing $d$ and thus the periodicity, $s+d$. The corresponding simulated reflectance spectra presented by the grey-dotted lines show an overall good agreement with respect to the experimental spectra. The simulated colours using the simulated reflectance spectra and the Commission Internationale de l'Éclairage (CIE) 1931 colour matching functions agree well with the observed colours. On the other hand, when $s$ increases successively ($d = 300$ nm), colour tuning is less dramatic (see the right plot in Fig. 1c). The reflectance peak experiences red-shifts (see the red-dashed line) with increasing $s$. Meanwhile, a new reflectance peak is observed at shorter wavelengths. The corresponding simulated spectra presented by the grey-dotted lines and the simulated colours both agree well with the experimental results. The details about spectral and colour simulations can be found in Methods.

To elucidate the underlying physics of the observed spectral characteristics, contour maps of the resonance positions in dependence on $d$ (see Supplementary Fig. S2a) and $s$ (see Supplementary Fig. S2b) were calculated, respectively. The contributions from the localized surface plasmon resonances (LSPRs)[22,23] of the individual Mg nanoparticles and the Rayleigh-Wood anomalies (RAs) of the particle arrays are indicated using black and white lines, respectively. As shown in Supplementary Figs. S2a and S2b, the longer-wavelength reflectance peak located in between the black and white-solid lines is associated with a hybrid plasmonic mode[24], which stems from the excitations of the LSPR and the RA for the $Mg/SiO_2$ interface. The RA for the air/Mg interface is presented by the white-dotted line.

To provide deeper insights into the mechanisms of the plasmonic modes, as a representative example, electric field and charge distributions at the respective resonance positions (indicated by the red arrows in the right plot in Fig. 1c) for the colour square ($s$150, $d$300) were calculated. At the resonance of 710 nm, the charge distribution reveals the excitation of a dipolar LSPR mode in the Mg nanoparticle (see Supplementary Fig. S3a). The corresponding electric field distribution demonstrates that the electric fields are confined at the corners of the particle near the $Mg/SiO_2$ interface. At the resonance of 430 nm, the charge distribution reveals the excitation of a quadrupolar LSPR mode (see Supplementary Fig. S3b). The associated electric fields are localized at the corners of the particle near the air/Mg interface.

The dynamic behaviour of the plasmonic pixels is enabled by the unique phase-transition of Mg in response to hydrogen[25–30]. Mg can absorb up to 7.6 wt % of hydrogen and subsequently undergoes a metal-to-dielectric transition to form magnesium hydride ($MgH_2$)[25]. This remarkable capacity surpasses all known reversible metal hydrides in hydrogen uptake capability. Upon hydrogen exposure, the Pd capping layer catalyzes the dissociation of hydrogen molecules into hydrogen atoms, which then diffuse through the Ti capping layer into the Mg particle[19,20,28]. The Ti adhesion and capping layers help to release the mechanical stress from volume expansions of Mg (32%) and Pd (11%). The Ti capping layer also plays an important role as spacer to prevent Mg and Pd from alloying[19,20]. A movie that records the colour dynamics of the palette upon 10% hydrogen exposure (in nitrogen carrier gas) is presented in Supplementary Movie 1a. Different colour squares undergo a series of vivid colour changes until all colours vanish. The hydrogenation process is rather complicated and essentially associated with a gradual decrease of the metallic fraction of the particles, formation of $MgH_2$ as dielectric surrounding, and simultaneous hybrid particle volume expansion. Such a catalytic process renders dynamic alterations to the reflectance spectra possible as shown in Supplementary Fig. S4a, thus leading to dynamic colour changes. After hydrogenation, the reflectance spectra become nearly featureless as observed from both the experimental and simulated reflectance spectra in Supplementary Fig. S4a and Fig. S5. When all the Mg particles are transformed into $MgH_2$ particles, the colours of the palette are erased. The optical images of the palette under different illumination intensities can be found in Supplementary Fig. S6. It is worth mentioning that the colour changing speed during the hydrogenation process can be manipulated by several key parameters including the thickness of the Pd catalytic layer, the hydrogen concentration, and the substrate temperature. In this work, 10 nm Pd, 10% hydrogen, and room temperature are employed in the different display demonstrations, if not particularly specified.

To explore the colour transformation kinetics during hydrogenation, evolutions of the exemplary colour squares are investigated as shown in Fig. 1d. The snapshot images were extracted from Supplementary Movie 1a and are arranged as a function of *ln(t)*. The top figure in Fig. 1d corresponds to the evolution of the colour squares, in which the constituent Mg particles have the same size but different interparticle distances. It illustrates that smaller interparticle distances are associated with richer colour changes over time, exhibiting a dynamic transition from yellow, red, blue, colour fading to complete colour erasing. In contrast, larger interparticle distances are essentially correlated with a blue colour diminishing process. Interestingly, the erasing times for different colour squares are nearly identical (see the vertical grey-dashed line). The bottom figure in Fig. 1d presents the evolution of the colour squares, in which the Mg particles have the same interparticle distance but different sizes. The erasing times differ significantly for different colour squares as indicated by the slanted grey-dashed line. Evidently, colour squares that contain smaller Mg particles vanish faster than those composed of larger particles. This elucidates clear size-dependent hydrogenation kinetics of the Mg nanoparticles.

Importantly, the erased colours can be restored through dehydrogenation of the Mg particles in the presence of oxygen[19,20]. Such reversible colour transformations are of great importance for dynamic display applications. A movie that records the colour recovering of the palette at a concentration of 20% oxygen (in nitrogen carrier gas) is presented in Supplementary Movie 1b. The dynamic reflectance spectra of the representative colour squares during dehydrogenation can be found in Supplementary Fig. S4b. The oxidative dehydrogenation process involves binding of oxygen with the desorbed hydrogen atoms from $MgH_2$ to form $H_2O$[29,30]. This avoids a build-up of hydrogen at the Pd surface, thus facilitating hydrogen desorption. As shown in Supplementary Movie 1b, the colours of the palette can be nicely restored. This indicates that the $MgH_2$ particles are converted back to Mg particles, therefore displaying brilliant colours again. The dynamic CIE maps for the palette and the selected colour squares in Fig. 1c during hydrogenation/dehydrogenation can be found in Supplementary Fig. S7 and Fig. S8, respectively.

Hysteresis is observed, when the colours of the palette evolve along the hydrogenation and dehydrogenation pathways, respectively. This is also seen from the recorded dynamic spectra of the selected colour squares as shown in Supplementary Fig. S4. The hysteresis behaviour can be attributed to the different changes of the particle geometries, the local hydrogen concentrations in the Mg particles, and the propagation directions of the $Mg/MgH_2$ interfaces during hydrogenation and dehydrogenation, respectively. A simple model has been proposed that after hydrogenation a layer of $MgH_2$ is formed directly beneath the hydrogenated Ti layer and it proceeds towards the substrate[19]. This results in successive spectral red-shifts due to effective thickness decreases of the Mg particles (see Supplementary Fig. S4a). Subsequently, during dehydrogenation the optical spectra exhibit spectral shifts with hysteresis (see Supplementary Fig. S4b). A systemic study on the complex hydrogenation and dehydrogenation processes in the Mg particles, for example, using environmental transmission electron microscopy and atomic force microscopy will be highly desirable for providing an insightful understanding of the local catalytic kinetics in the Mg particles and their correlated dynamic optical spectra.

Such plasmonic colour tuning, erasing, and restoring based on catalytic metasurfaces open an avenue for a variety of dynamic colour display applications. A direct application is to produce high-quality colour-tunable plasmonic microprints. As a demonstration, the Minerva logo of the Max-Planck Society has been prepared using selected colours and their corresponding matrix numbers from the palette are indicated in Fig. 2a. A description about the layout generation approach can be found in Supplementary Fig. S9. A discussion about the NA dependence and the angle dependence of the colours can be found in Supplementary Fig. S10. The representative snapshot images of the selected colour squares during hydrogenation and dehydrogenation are shown in Fig. 2a. The SEM images illustrating the details of the Minerva logo are presented in Fig. 2b. Fig. 2c shows the performance of the dynamic plasmonic display, and the accompanying movie can be found in Supplementary Movie 2a. Upon hydrogen loading, the Minerva logo undergoes dynamic colour changes with the individual colour blocks following the changing routes of the respective colour squares (see Fig. 2a). A series of abrupt colour alterations take place within 23 s. Subsequently, the logo starts to fade and completely vanishes after 566 s. Owing to the identical sizing of the constitute Mg nanoparticles, different colours in the logo are erased nearly simultaneously, as discussed in Fig. 1d. Upon oxygen exposure, the logo is gradually restored to its starting state without experiencing drastic colour changes (see Supplementary Movie 2b). The restoring process takes approximately 2,224 s to complete. To enhance the durability of our samples, exhaustive optimization of the fabrication procedures has been carried out (see Methods). As a demonstration of excellent reversibility and durability, operation of the Minerva logo display in a number of cycles is shown in Supplementary Movie 2c. In addition, operation of a Heidelberg University logo display at elevated temperatures in more cycles is shown in Supplementary Movie 2d.

An interesting functionality of our display technique is image 'freezing' at any designated display state, owing to the unique hydrogenation process of Mg, which can be paused in between a fully metallic state (Mg) and a fully dielectric state ($MgH_2$). To demonstrate this aspect, in a new cycle of hydrogenation as shown in Fig. 2d, the logo display is held at an arbitrary state by switching off hydrogen. For example at 11 s, hydrogen is switched off and only pure nitrogen is present. As a result, the dynamic process halts immediately and the logo display is 'frozen' at the specific colour state (see the green box in Fig. 2d). When hydrogen is switched on again, the dynamic process proceeds subsequently. The corresponding Movie is presented in the second cycle in Supplementary Movie 2c.

Remarkably, the distinct colour transformation kinetics of the Mg particles allows for the realization of the first plasmonic animations with subwavelength resolution. To demonstrate this functionality, a variety of fireworks have been designed as shown in Fig. 3a. The matrix numbers of the selected colours from the palette are labeled next to the individual firework schemes. One of the underlying design mechanisms lies in the fact that Mg particles of different sizes but identical interparticle distance are subject to dramatically distinct colour erasing times during hydrogenation. To this end, fireworks I and IV are designed to display radially explosive effects. Firework V is a spirally propagating animation. Firework II combines both the radial and spiral effects. On the contrary, the inner part of firework III essentially employs Mg particles of identical size but different interparticle distances, which allow for a propagating animation with rich colour variations. The outermost part of firework III utilizes the same design principle as firework I. Fig. 3b presents the optical microscopy image of the fabricated plasmonic firework display. A Movie that records the firework animations upon hydrogen exposure is shown in Supplementary Movie 3. The representative snapshot images of the fireworks during hydrogenation are presented in Fig. 3c. All the fireworks demonstrate dynamic animation effects, which agree well with our design. For example, the balloon firework I displays a clear radially explosive effect, dispatching wave-like colour changes from yellow to orange along the radial axis. Afterwards, the orange colour diminishes along the backward direction until the firework completely disappears. The spiral firework V exhibits a vivid propagating effect following a spiral route with the blue colour gradually disappearing from the exterior.

Importantly, our dynamic plasmonic display technique is ideally suited for highly secure information encryption and anti-counterfeiting applications[4,15]. To achieve a superior security level, a fabrication protocol with smart material processing has been developed. The fabrication procedures are illustrated in Fig. 4a. The first information pattern (for example, a letter 'N') is defined in a PMMA resist using electron-beam lithography (EBL), followed by Ti/Mg/Ti (3 nm/50 nm/5 nm) evaporation using an electron-gun evaporator. Subsequently, the sample is placed in ambient air for 10 min and then loaded back to the evaporator for Pd (10 nm) evaporation. After lift-off, the second information pattern (for example, a letter 'A') is created using EBL with alignment markers, followed by Ti/Mg/Ti/Pd (3 nm/50 nm/5 nm/10 nm) evaporation and lift-off. Due to the formation of thin oxide layers between Pd and Mg, the letter 'N' is inactive to hydrogen. On the contrary, the letter 'A' is active to hydrogen. As shown by the SEM images in Fig. 4b, the letters in each line do not exhibit any discernable differences, neither can the encrypted information be differentiated using optical microscopy (see the left image in Fig. 4c). Only upon hydrogen exposure, the hidden code 'nano Printing' is decrypted as shown in the right image in Fig. 4d. The information decryption takes approximately 10 min to complete. The accompanying movie is presented in Supplementary Movie 4a. In this regard, information encoded on selected pixels, which are indiscernible to both optical and scanning electron microscopies, can only be read out using hydrogen as decoding key, demonstrating the superior security level of our display technique. Importantly, the information can be reversibly encrypted using oxygen within 40 min (see Supplementary Movie 4b). The predicted display images before and after hydrogenation are shown in Supplementary Fig. S11a. The associated colour evolutions for the decryption and encryption processes are shown in Supplementary Fig. S11b.

Crucial for practical applications, our technique can also be used to create arbitrary dynamic microprints with excellent colour and tonal control. To demonstrate this ability, Vincent van Gogh's *Flowers in a Blue Vase* has been utilized as design blueprint (see Fig. 5a left). To particularly highlight the colour dynamics of the flowers, the background of the original artwork was removed. The modified design and the optical microscopy image of the fabricated artwork display are presented in Fig. 5a. The corresponding SEM images are shown in Fig. 5b. Fig. 5c presents the colour erasing and restoring of the dynamic artwork display. The accompanying movie can be found in Supplementary Movie 5.

## Discussion

In summary, we have presented a novel dynamic plasmonic display technique based on catalytic Mg metasurfaces. The excellent plasmonic properties and unique hydrogenation/dehydrogenation kinetics afforded by Mg nanoparticles enable dynamic plasmonic displays with unprecedented functionalities and subwavelength resolution. Careful material engineering and optimization using Mg alloys[25–27] can be carried out to further improve the display durability for real world applications. In addition, polytetrafluoroethylene protective coating can be applied to avoid water staining on the display surface[31,32]. Our technique suggests promising avenues for applications in actively tunable displays[8–11] and filters[7–9,33], plasmonic holograms[34,35], plasmonic colourimetric sensing[36], advanced optical data storage[14], security tagging and cryptography[15], as well as realization of plasmonic movies with subwavelength resolution in the future.

## Methods

**Fabrication of Mg metasurfaces.** In order to enhance the durability of the samples, exhaustive optimization of the fabrication procedures was carried out. First, a double-layer PMMA resist (250k-2.5% and 950k-1.5%, Allresist) was spin-coated (5 s at 3,000 rpm and 30 s at 8,000 rpm, respectively) on a $SiO_2$ (100 nm)/Si substrate, followed by baking (4 min at 160 ºC) on a hotplate. Electron-beam lithography (EBL, Raith eLine) was performed with 20 kV acceleration voltage, 20 μm aperture, 130 pA beam current, 6.4 nm exposure step size, 500 μC·cm$^{-2}$ dose, and 100 × 100 μm$^2$ writing field. After development (90 s in MIBK and 60 s in isopropanol), an oxygen plasma treatment (5 s) was employed to clean the PMMA opening areas. 3 nm Ti, 50 nm Mg, 5 nm Ti, and 10 nm Pd were successively deposited on the substrate through electron-gun evaporation (PFEIFFER Vacuum, PLS-500). The presence of the Ti adhesion layer was found to result in significantly improved particle morphology. The deposition rates for Mg, Ti, and Pd were 1.0, 0.1, and 0.2 nm/s. Instead of following a standard routine using n-methyl pyrrolidinone (NMP) at an elevated temperature of 65 °C, metal lift-off was carried out in acetone at room temperature, which was found to help preserving the catalytic activity of the Mg nanoparticles over time very well.

**Optical measurements.** The colour images were revealed using a NT&C bright-field reflection microscopy setup (using a Nikon ECLIPSE LV100ND microscope) illuminated by a white light source (Energetiq Laser-Driven Light Source, EQ-99). A digital CCD Camera (Allied-Vision Prosilica GT2450C) was used to capture the colour micrographs with a 20 × (NA = 0.4) objective. The optical spectra were measured in reflection mode using a microspectrometer (Princeton Instruments, Acton SP-2356 Spectrograph with Pixis:256E silicon CCD camera) with the electric field of the unpolarized light in plane with the substrate surface. The measured reflectance spectra were normalized with respect to that of a bare substrate. All measurements were carried out at room temperature (~ 25 ºC). The flow rate of hydrogen and oxygen was 2.0 L/min.

**Numerical simulations.** Numerical simulations were carried out using commercial software COMSOL Multiphysics based on a finite element method (FEM). Periodic boundary conditions and waveguide port boundary conditions were used for calculation of the structure arrays. Perfectly matched layers and background field conditions were used for calculations of the single structures. The simulations were carried out with the substrate. The refractive index of $SiO_2$ was taken as 1.5. The dielectric constants of Si, Mg, and $MgH_2$ were taken from Green[37], Palik[38], and Griessen[39]. The dielectric constants of Pd and PdH were taken from Ref. 40.

**Chromaticity calculations.** The colours of different colour squares were calculated using the simulated reflectance spectra and the colour-matching functions defined by the International Commission on Illumination (Commission Internationale de l'Éclairage, CIE).[41] The spectral power distribution is given by:

$$P(\lambda) = I(\lambda)R(\lambda) \quad (1)$$

$I(\lambda)$ is the relative radiance spectrum of the white light source. $R(\lambda)$ is the simulated reflectance spectrum. The tristimulus values *X*, *Y*, and *Z* are computed through:

$$X = \frac{1}{K}\int_\lambda \bar{x}(\lambda)P(\lambda)d\lambda \quad (2a)$$

$$Y = \frac{1}{K} \int_\lambda \bar{y}(\lambda) P(\lambda) d\lambda \tag{2b}$$

$$Z = \frac{1}{K} \int_\lambda \bar{z}(\lambda) P(\lambda) d\lambda \tag{2c}$$

Here, $\bar{x}(\lambda)$, $\bar{y}(\lambda)$, and $\bar{z}(\lambda)$ are the CIE standard observer functions. The integrals are computed over the visible range (from 380 nm to 780 nm). $K$ is the normalizing constant:

$$K = \int_\lambda \bar{y}(\lambda) I(\lambda) d\lambda \tag{3}$$

The CIE chromaticity coordinate $(x, y, z)$ can be obtained by normalization:

$$x = \frac{X}{X + Y + Z} \tag{4a}$$

$$y = \frac{Y}{X + Y + Z} \tag{4a}$$

$$z = \frac{Z}{X + Y + Z} = 1 - x - y \tag{4a}$$

It is noteworthy that only two values of $(x, y, z)$ are independent, because the intensity of the incident light source is normalized. To fit the real illumination condition, we transform the coordinate $(x, y, z)$ into the HSL $(h, s, l)$ colour space. HSL is the abbreviation for Hue, Saturation, and Luminance. Hue specifies the base colour. Luminance corresponds to the intensity of the reflected light, and it can be adjusted based on the experiment to calibrate the chromaticity calculation conveniently.

**Corresponding Author**
*E-mail: laura.liu@is.mpg.de

**Author Contributions**
X.D. and N.L. conceived the project. X.D. performed the experiments and theoretical calculations. X.D. developed the codes. S.K. built the gas system, and helped with the Mg material processing.

**Acknowledgements**
We gratefully acknowledge the generous support by the Max-Planck Institute for Solid State Research for the usage of clean room facilities. We also thank the 4th Physics Institute at the University of Stuttgart for kind permission to use their electron-gun evaporation system. We thank Thomas Weiss for helpful discussions and Yvonne Link for initial tests of Mg evaporation. This project was supported by the Sofja Kovalevskaja grant from the Alexander von Humboldt-Foundation, the Marie Curie CIG grant, and the European Research Council (ERC *Dynamic Nano*) grant.

**Additional information**
Supporting Information is available in the online version of the paper.

**Competing financial interests**
The authors declare no competing financial interest.

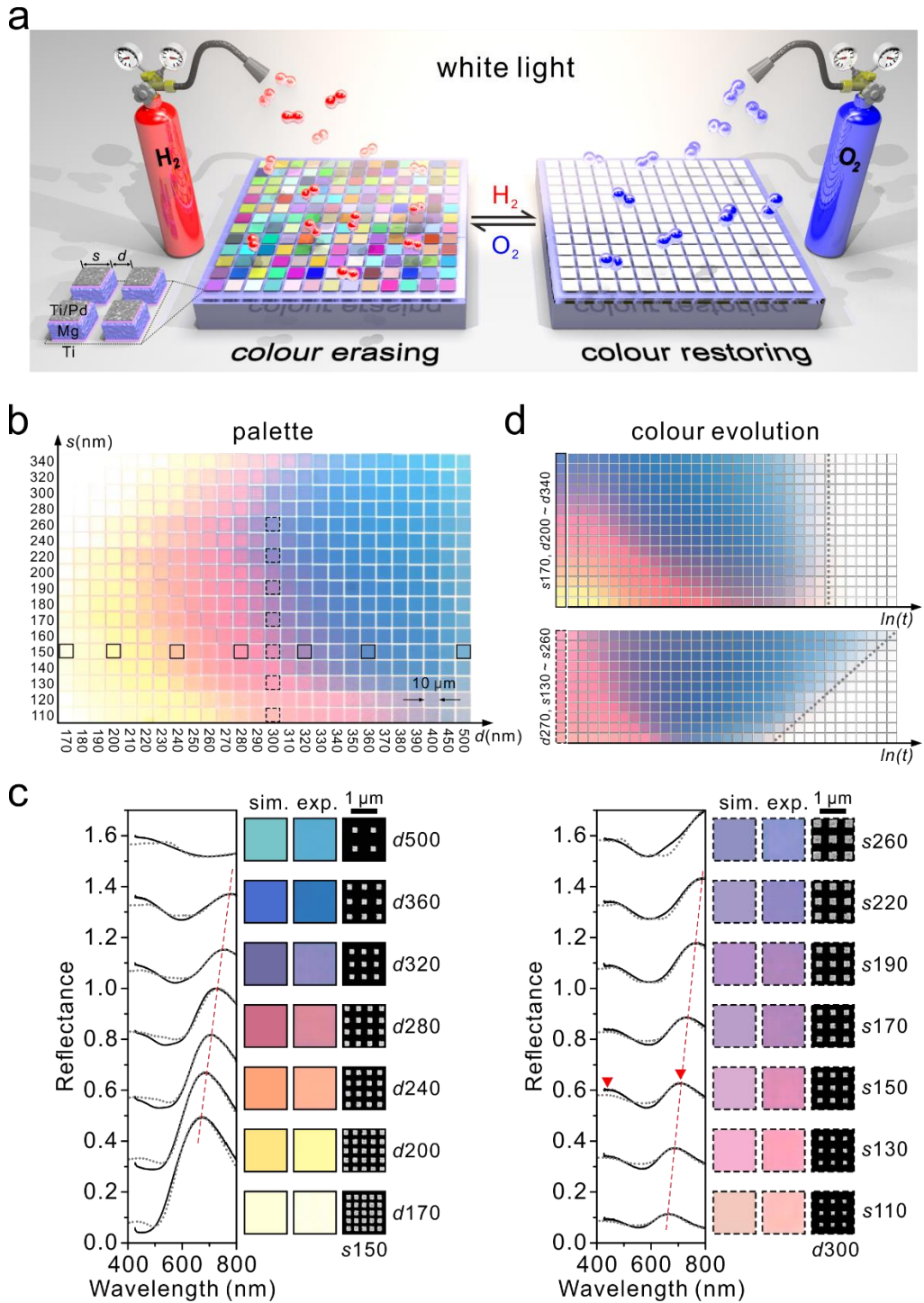

**Figure 1 | Working principle and optical characterization of the dynamic plasmonic colour display. a**, Schematic of the plasmonic metasurface composed of hydrogen-responsive Mg nanoparticles interacting with incident unpolarized white light. These Mg nanoparticles are sandwiched between Ti/Pd capping layers and a Ti adhesion layer. In each colour square (10 μm), they are arranged in a lattice with a period of $s+d$ along both directions. $s$ and $d$ are the size of the square-shaped particles and the interparticle distance, respectively. The colours of the plasmonic metasurface can be erased upon hydrogen exposure through phase-transition from Mg to MgH$_2$. The colours can be restored upon oxygen exposure through transformation of MgH$_2$ back to Mg. **b**, Colour palette

obtained by stepwise tuning of *s* and *d*. **c**, Experimental (black) and simulated (grey-dotted) reflectance spectra of the colour squares selected from **b**. The spectral curves are shifted upwards for clarity. In the left (right) panel, the selected colours correspond to those indicated using solid (dashed) squares along the horizontal (vertical) direction in **b**. Experimental and simulated colours as well as the corresponding SEM images of the structures. The red dashed line indicates the shift of the reflectance peak. **d**, Colour evolutions of the selected colour squares upon hydrogen exposure over time *ln(t)*. The grey-dotted lines indicate the colour vanishing times in the two cases.

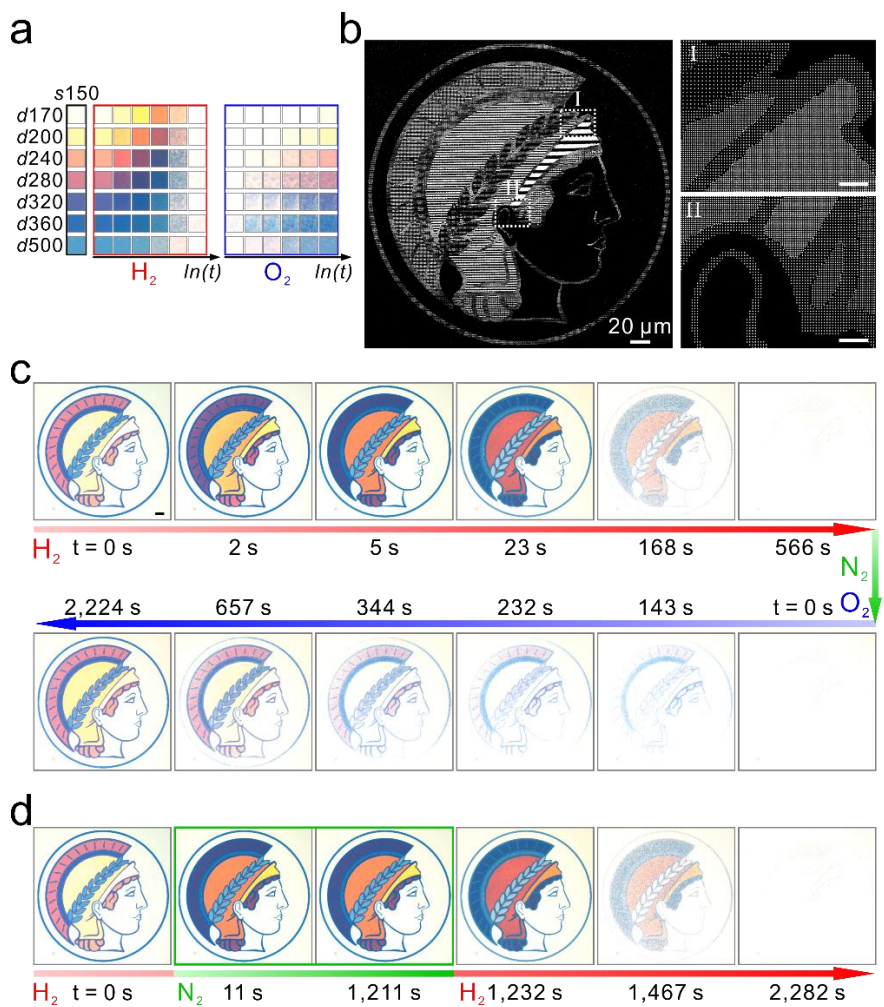

**Figure 2 | Dynamic colour tuning, erasing, and restoring of the Max-Planck-Society's Minerva logo microprint. a,** Colour evolutions of the selected colours used to construct the Minerva logo during hydrogenation and dehydrogenation, respectively. **b**, Overview and enlarged SEM images of the Minerva logo. The scale bars in I and II are 5 μm. **c**, Optical micrographs of the Minerva logo during hydrogenation and dehydrogenation for colour erasing and restoring, respectively. Scale bar: 20 μm. **d**, Colour state 'freezing' by switching off hydrogen.

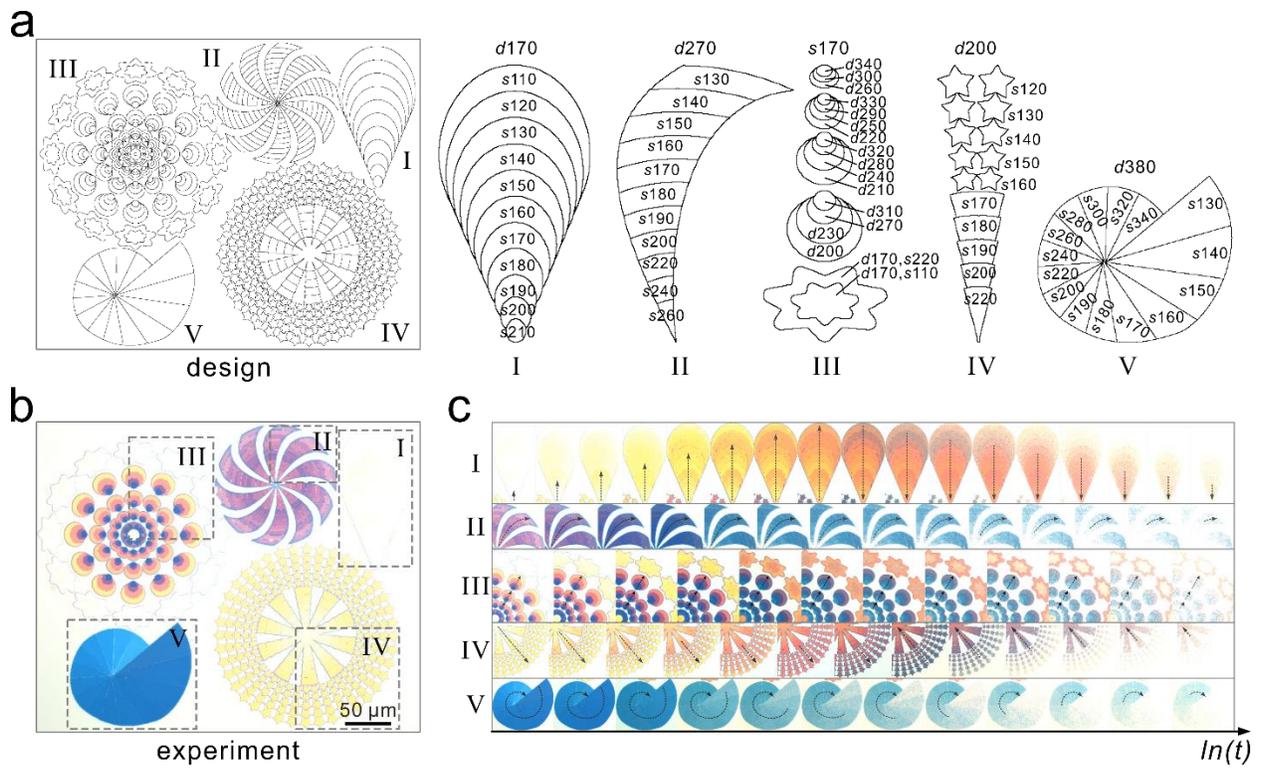

**Figure 3 | Plasmonic subwavelength animations. a,** Schematics of the firework animations characterized with different dynamic effects. **b,** Optical micrograph of the fireworks. **c,** Selected snapshots of the firework animations over time $ln(t)$, illustrating the individual dynamic effects.

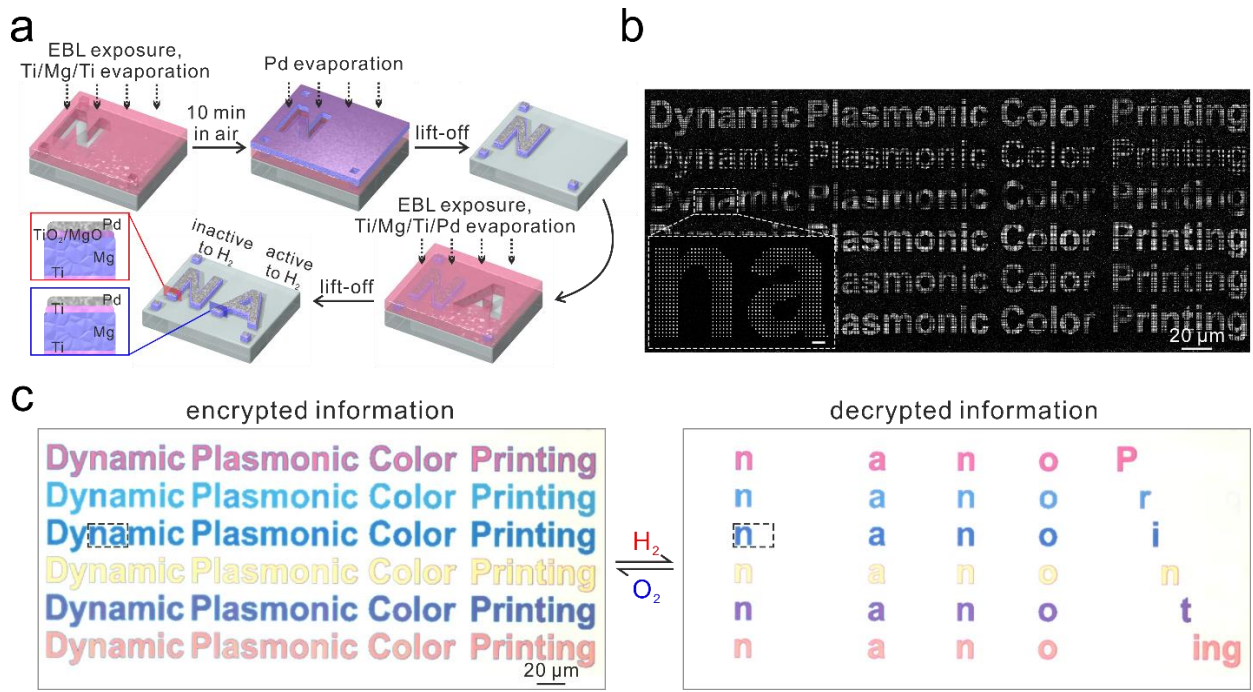

**Figure 4 | Highly secure information encryption. a,** Schematic of the fabrication process. **b,** Overview and enlarged SEM images of the encrypted microprint. Inset scale bar: 2 μm. **c,** Optical micrographs of the encrypted plasmonic display (left) and the decrypted plasmonic display (right).

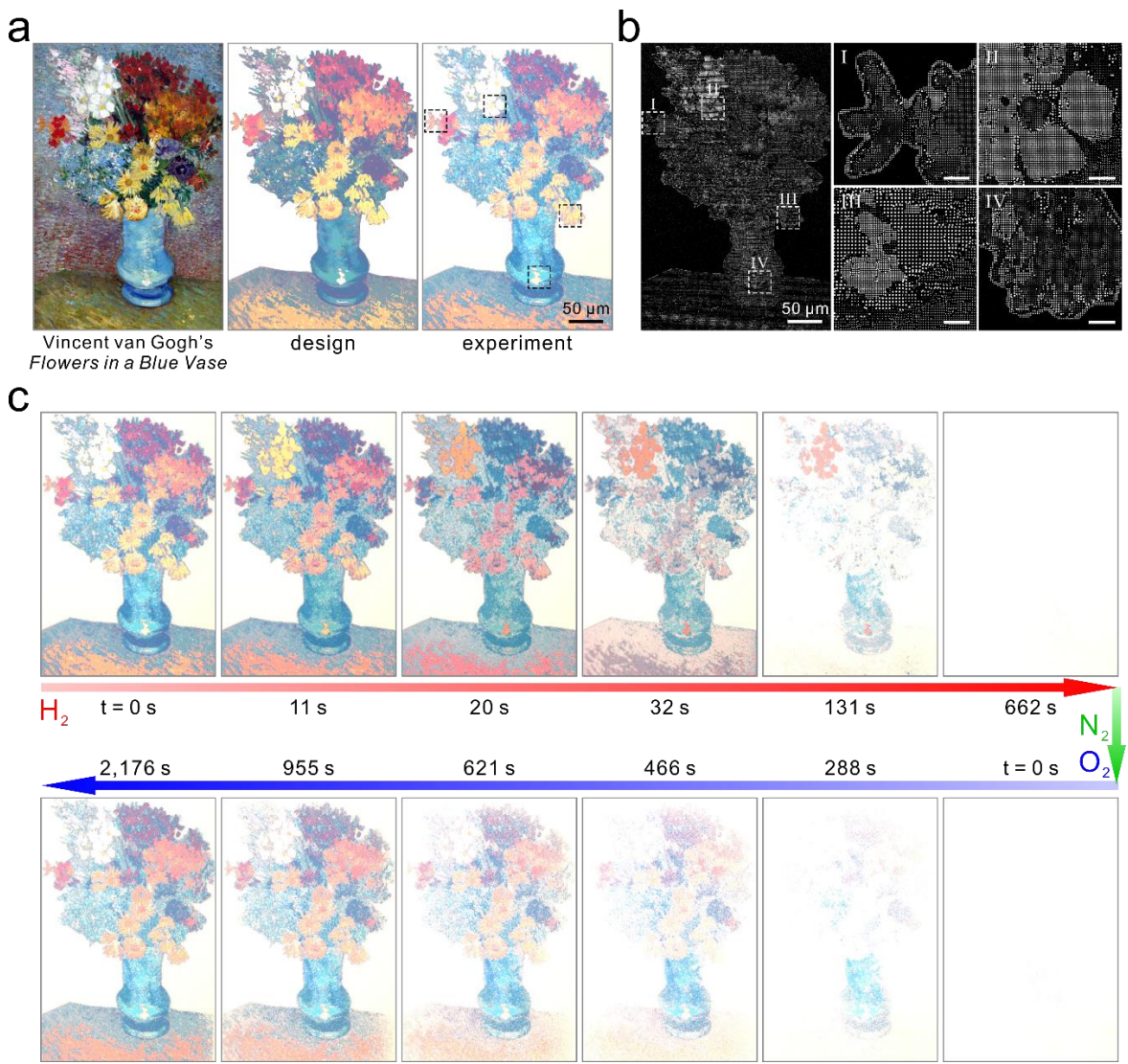

**Figure 5 | Dynamic display of arbitrary images. a,** Left: Digital copy of a Vincent van Gogh painting (*Flowers in a Blue Vase*). Middle: Modified design of the original artwork in order to highlight the dynamic changes of the flowers. Right: Optical micrograph of the modified artwork. **b,** Overview and enlarged SEM images of the microprint. Scale bars are 5 µm in the enlarged images. **c,** Dynamic processes of the plasmonic artwork display during hydrogenation and dehydrogenation for colour erasing and restoring, respectively.